\documentclass[12pt,preprint]{aastex}
\begin{document}
\title{A Statistical Strategy for the Sunyaev-Zel'dovich Effects Cluster Data}
\author{Jounghun Lee}
\affil{Department of Physics, University of Tokyo,Tokyo 113-0033, Japan}
\email{lee@utap.phys.s.u-tokyo.ac.jp}

\begin{abstract} 

We present a statistical strategy for the efficient determination 
of the cluster luminosity function from the interferometric 
Sunyaev-Zel'dovich (SZ) effects cluster data.  
To determine the cluster luminosity function from the noise contaminated 
SZ map, we first define the zeroth-order cluster luminosity function 
as difference between the measured peak number density of the SZ map 
and the mean number density of noise. Then we demonstrate that the noise 
contamination effects can be removed by the stabilized deconvolution of 
the zeroth-order cluster luminosity function with the one-dimensional 
Gaussian distribution.  We test this analysis technique against Monte-Carlo 
simulations, and find that it works quite well especially in the 
medium amplitude range where the conventional cluster selection 
method based on the threshold cut-off usually fails. 

\end{abstract} 

\newcommand{\etal}{{\it et al.}}
\newcommand{\beq}{\begin{equation}}
\newcommand{\eeq}{\end{equation}}
\newcommand{\ben}{\begin{eqnarray}}
\newcommand{\een}{\end{eqnarray}}
\newcommand{\bS}{{\bf S}}
\newcommand{\bL}{{\bf L}}
\newcommand{\bth}{\vec{\theta}}
\newcommand{\hbth}{\hat{\theta}}
\newcommand{\bu}{{\bf u}}

\section{INTRODUCTION}

Galaxy clusters are the biggest bound objects of the universe.  
Being rare and formed relatively late, the abundance of galaxy 
clusters depends sensitively on the background cosmology 
\citep{bar-etal96,hen97,bah-fan98,via-lid99,
bor-etal99,fan-chi01,gre-etal01,mol-etal02}. 
To find as many galaxy clusters as possible and to investigate 
the evolution of their abundance have thus become two of the most 
challenging tasks of the current observational cosmology.       
The cluster survey is categorized by the observing waveband as  
the X-ray, the optical, and the Sunyaev-Zel'dovich effect.  
In the past the X-ray or the optical surveys were favored due 
to their high rate of detecting clusters and relatively low-costs. 
However, thanks to recent development in technology, 
the Sunyaev-Zel'dovich effect is currently spotlighted as a powerful  
cosmological probe \citep{car-etal96,whi-etal99,hol-etal00,lo-etal00}
that can provide a statistically unbiased sample of clusters unlike 
the X-ray and the optical surveys.  

The Sunyaev-Zel'dovich (SZ) effect represents the change in brightness 
of the cosmic microwave background (CMB) radiation caused by the
interaction of the CMB photons with the ionized intra-cluster gas 
\cite{sun-zel72}.  The underlying physics of the SZ effect is the 
inverse Compton scattering: free electrons of the hot intra-cluster 
gas scatter off the CMB photons as they pass through the galaxy 
clusters, which results in the shift of the CMB photon frequency 
and a corresponding change in its radiation energy. 
Depending on whether the scattering of the CMB photons was by 
the random or systematic motions of the electrons, the SZ effect 
is called thermal or kinetic respectively. Here we focus on the 
thermal SZ effect since its contribution is order of magnitude 
stronger than the kinetic counterpart \citep{spr-etal01,zha-etal02}.
 
The SZ effects ($\Delta I_{f}$) observed at a given frequency 
($f$) at a given position $\bth$ on the sky depend on the 
electron number density ($n_{e}$)  and temperature ($T_{e}$) within 
the galaxy clusters that the CMB photons have encountered all along 
the path:       
\begin{equation}
\Delta I_{f}(\bth) = -2g_{f}\frac{\sigma_{T}k_{B}}{m_{e}c^{2}}
\int n_{e}(l\hbth)T_{e}(l\hbth)dl \equiv -2g_{f}y(\bth), 
\label{eqn:thomson} 
\end{equation} 
with the integral taken along the CMB path in the line-of-sight 
direction ($\hbth \equiv \bth/\vert \bth \vert$).
Here $y$ is the cluster Comptonization parameter, and  
$g_{f}$ is the frequency-dependent spectral shape factor.  
Note that the SZ effects are spatially localized with the associated 
clusters along the line of sight, unlike the primordial CMB
fluctuations.

Equation (\ref{eqn:thomson}) implies that the amplitude of the SZ 
effects observed in a narrow frequency band with almost fixed $g_{f}$ 
can be quantified by the $y$-parameter. Provided that noise were 
absent in the SZ map, clusters could be counted as local maxima of $y$.    
In practice, however, noise always dominate the SZ map.  
A critical issue is how to eliminate the dominant noise-contamination 
effects.  A common observational practice to identify true signals from 
the noise-contaminated field is to select only such peaks with amplitudes 
above some threshold (usually several times the noise standard deviation). 
Using this technique, however, one can select only high-amplitude 
signals for a given observational integration time. In order to 
increase the signal to noise ratio, one has to decrease the noise 
level by increasing the integration time, which is proportional only 
to the square of the the noise level.  

Considering the usual high-costs of the SZ experiments, this 
conventional technique of signal selection is too inefficient 
to apply to the SZ cluster data .  One may wish to find a more 
efficient statistical strategy that can allow us to determine the cluster 
luminosity function as quickly and accurately as possible.  
Powerful new generations of SZ instruments such as AMiBA 
(Array for Microwave Background Anisotropy, see Lo \etal\ 2001)  
are already in the pipeline.  Expecting plenty of cluster data 
coming out in a few years, it is quite urgent in fact to develop such 
statistical strategies now.  

In this Letter we develop a statistical strategy for the efficient 
determination of the cluster luminosity function, 
especially from interferometric SZ surveys using AMiBA as our model 
experiment.

\section{SZ CLUSTER LUMINOSITY FUNCTION }

The SZ cluster luminosity function, $n_{cl}(y)$, is defined as the 
number density of galaxy clusters with the associated cluster Compton 
parameter in the range of $[y,y+dy]$. Its cumulative function,
$N_{cl}(\ge y) \equiv \int _{y}n_{cl}(y^{\prime}) dy^{\prime}$, is 
connected to the cluster mass function  (see eq.[7] in Barbosa \etal\
1996), and thus can be used in principle as a cosmological discriminator 
\citep{hol-etal00,fan-chi01,die-etal02,gre-etal01,mol-etal02,ben-etal02}. 
In practice, the direct conversion of the cluster luminosity 
function into the cluster mass function is fraught with the 
difficulties related to the rather large scatter in the correlation
between cluster mass and the SZ effect strength  
\citep[e.g., see][]{met02}.   

Anyway, our goal here is to find an efficient way to find $n_{cl}(y)$ 
from the observed SZ map that is expected to be significantly 
contaminated by noise.  There are two different sources of noise: 
instrumental noise and primordial CMB fluctuations. 
The primordial fluctuations, however, turned out to be negligible in 
interferometric SZ cluster surveys \cite{zha-etal02}. 
In the following analysis, we concentrate on instrumental noise only.  

Our model experiment, AMiBA, as an interferometric SZ survey, will 
employ the drift-scan method to optimize the observations. For the 
detailed description of AMiBA and the drift-scan method, see   
Lo \etal\ (2000) and Pen \etal\ (2002) respectively. 
Among many advantages of the drift-scan method, it makes noise-analysis 
tractable: in a SZ map measured from the drift-scanned CMB sky, 
noise is Gaussian white. Therefore, a total SZ map measured by AMiBA
will be a combination of non-Gaussian cluster sources with Gaussian
white.  In the following two subsections, 
we simulate a total SZ map by means of the Monte-Carlo method, 
and reconstruct the cluster luminosity function by eliminating 
the noise contamination effects from the SZ map.  

\subsection{Monte-Carlo Simulations of Drift-Scan SZ Maps}

We have constructed a random field on a $2048^{2}$ mesh in a periodic 
box of linear size  $1$ deg using the Monte-Carlo method, 
in such a way that the random field possesses the main statistical 
properties of a cleaned SZ map expected from AMiBA drift-scan survey 
over a unit area per a unit hour, assuming the flat-sky approximation. 
 
By a cleaned SZ map, we mean the SZ map smoothed by a optimal filter: 
In Fourier $\bu$-space  
($\bu$: the Fourier counterpart of $\bth$, $u \equiv \vert \bu \vert$), 
the optimal filter for AMiBA drift-scan observations is given 
\cite{pen-etal02} as  $W_{C}(u) = W_{F}(u)W_{N}(u)$, where 
$W_{F}(u)$ and $W_{N}(u)$ are the cluster intrinsic shape and the 
natural beam respectively.  
We use $W_{F}(u) = \frac{1}{u}$ and approximate $W_{N}(u)$ 
as $W_{N}(u) \approx \exp\left(-\frac{u^{2}\theta^{2}_{A}}{2}\right) 
- \exp\left(-\frac{u^{2}\theta^{2}_{B}}{2}\right)$, where  the 
two angular scales,  $\theta_{A}$ and $\theta_{B}$, represent 
the size of the natural and primary beam respectively, related 
to each other by $\frac{\theta_{A}}{\theta_{B}} = \frac{1}{6}$ 
(Pen 2002, private communications). 
We first constructed a Gaussian random field with the white noise 
power spectrum and convolved it by $W_{C}(u)$, and rescaled 
the field by its rms fluctuations, $\sigma^{2}_{0} = 
\frac{1}{(2\pi)^{2}}\int \vert W_{C}(u)\vert^{2}d^{2}{\bf u}$. 

Second, we simulated the cluster sources by generating a sparse set  
of two dimensional Gaussian functions of which peak locations and 
amplitudes were chosen randomly. Zhang \etal\ (2002) showed that 
the optimal scan rate for the purpose of AMiBA cluster search 
is around $150$ hours per square degree, which could find 1 cluster 
every $8$ hours.  Thus, the total number of the cluster sources was set 
to be $20$, with the expectation that the number of clusters per
square degree from the AMiBA cluster search would be around the same
number.  We locate the cluster sources in the two-dimensional map 
deliberately so that they are not overlapping one another.  
The size of each cluster source, i.e., the length scale of each
Gaussian function was chosen to be twice the pixel size, which is 
consistent with the expected AMiBA drift-scan map. 
The randomly chosen amplitudes of the cluster sources were in the 
range of $[0,5\sigma_{0}]$ where we would like to reconstruct the 
cluster luminosity function from the noise-contaminated map.  
The cluster amplitudes were chosen to be distributed exponentially, 
mimicking the real cluster distribution in this range 
\citep{zha-etal02}. 

Finally, we obtained a simulated SZ map by combining the Gaussian noise 
field with the cluster sources. Then we identified the local maxima of 
the total field by selecting those pixels whose amplitudes 
exceed the amplitudes of their $8$ closest-neighboring points, 
and count the number density of the total SZ map, $n_{sz}(\nu)$, 
as a function of the rescaled amplitude, $\nu \equiv y/\sigma_{0}$.

\subsection{Deconvolution Method}

If there were no clusters, the measured SZ map would be just 
a map of Gaussian noise, whose mean number density of local maxima, 
$n_{g}(\nu)$, is analytically derived to be \cite{lon57,bon-efs87}: 
\beq
n_{g}(\nu) = \frac{1}{(2\pi)^{3/2}}R_{*}^{-2}e^{-\frac{\nu^2}{2}}
\int^{\infty}_{0} [x^2+e^{-x^2}-1]
\frac{\exp[-\frac{1}{2}(x-\gamma\nu)^{2}/(1-\gamma^{2})]}
{[2\pi(1-\gamma^2)]^{1/2}}dx. 
\label{eqn:noi}
\eeq 
Here  
$\sigma^{2}_{i} = \frac{1}{(2\pi)^{2}}\int \vert
W_{C}(u)\vert^{2}u^{2i}d^{2}{\bf u}$ for our noise spectrum, and  
$R_{*} \equiv \sqrt{2}(\sigma_{1}/\sigma_{2})$, 
$\gamma \equiv \sigma_{1}^{2}/(\sigma_{0}\sigma_{2})$. 

The difference between the peak number density of the total SZ map 
and equation (\ref{eqn:noi}) provides a zeroth-order approximation 
to the cluster luminosity function: 
$n_{cl}^{(0)}(\nu) = n_{sz}(\nu) - n_{g}(\nu)$.  
At high $\nu$-tail ($\nu \gg 1$) where the mean number density 
of noise peaks drops practically to zero, 
$n_{sz}(\nu) \approx n_{cl}^{(0)}(\nu) \approx n_{cl}(\nu)$.  
At low-$\nu$ range ($0 \le \nu \le 2$) where Gaussian noise strongly 
dominates, $n_{sz}(\nu) \approx n_{g}(\nu)$ and $n_{cl}^{(0)}(\nu)$ 
measures just Poissonian noise scatter around equation (2).

The intriguing section is at the medium-$\nu$ range ($\nu \sim 3$),  
where $n_{cl}^{(0)}(\nu)$ includes not only the noise-scatter but 
also the noise-contaminated cluster peaks to a non-negligible degree. 
With the conventional cluster selection procedure based on the  
amplitude cut-off, the number density of cluster peaks in this
intermediate range cannot be determined since the peaks in this 
range are all disregarded as noise. Thus, it is this medium-amplitude 
section where one needs a better analysis technique to find the 
number density.   

Noise contaminates the cluster number density by changing its amplitude. 
Let $x$ measure the noise-contamination effects on the peak amplitude 
such that $\nu = \nu_{0} + x$ where $\nu_{0}$ and  $\nu$ are the real 
and contaminated amplitude of a cluster peak respectively.  
For a Gaussian noise,  $x$ can be assumed to be a Gaussian variable. 
Now, the probability distribution of $\nu$ can be written as a convolution 
of the probability distributions of $\nu_{0}$ and $x$.  
In terms of the number density, one can say 
\beq 
n_{cl}^{(0)}(\nu) = \frac{N_{cl}^{\rm tot}}{N_{sz}^{\rm tot}}
\int p(x)n_{cl}(\nu-x)dx,
\label{eqn:dec}
\eeq  
where $p(x)=\frac{1}{\sqrt{2\pi}}e^{-\frac{x^{2}}{2}}$ and 
$N_{sz}^{\rm tot}$ and $N_{cl}^{\rm tot}$ are the total number 
of peaks of the SZ map and that of real clusters respectively. 

Equation (\ref{eqn:dec}) implies that the cluster luminosity function, 
$n_{cl}(\nu)$, can be found by the deconvolution of the 
$n_{cl}^{(0)}(\nu)$  and $p(x)$.  In theory, the deconvolution of 
$n_{cl}^{(0)}(\nu)$ and $p(x)$ could be easily conducted : 
just dividing the Fourier transform of $n_{cl}^{(0)}(\nu)$ by that of 
$p(x)$, performing its inverse Fourier transformation. 
In practice, however, the deconvolution process itself is very 
unstable \cite{pre-etal92}.  To make the deconvolution stable 
and optimize the estimation of $n_{cl}(\nu)$, we apply a Wiener 
filter, say, $W_{P}(\nu)$, to $n_{cl}^{(0)}(\nu)$.  

\begin{figure}
\plotone{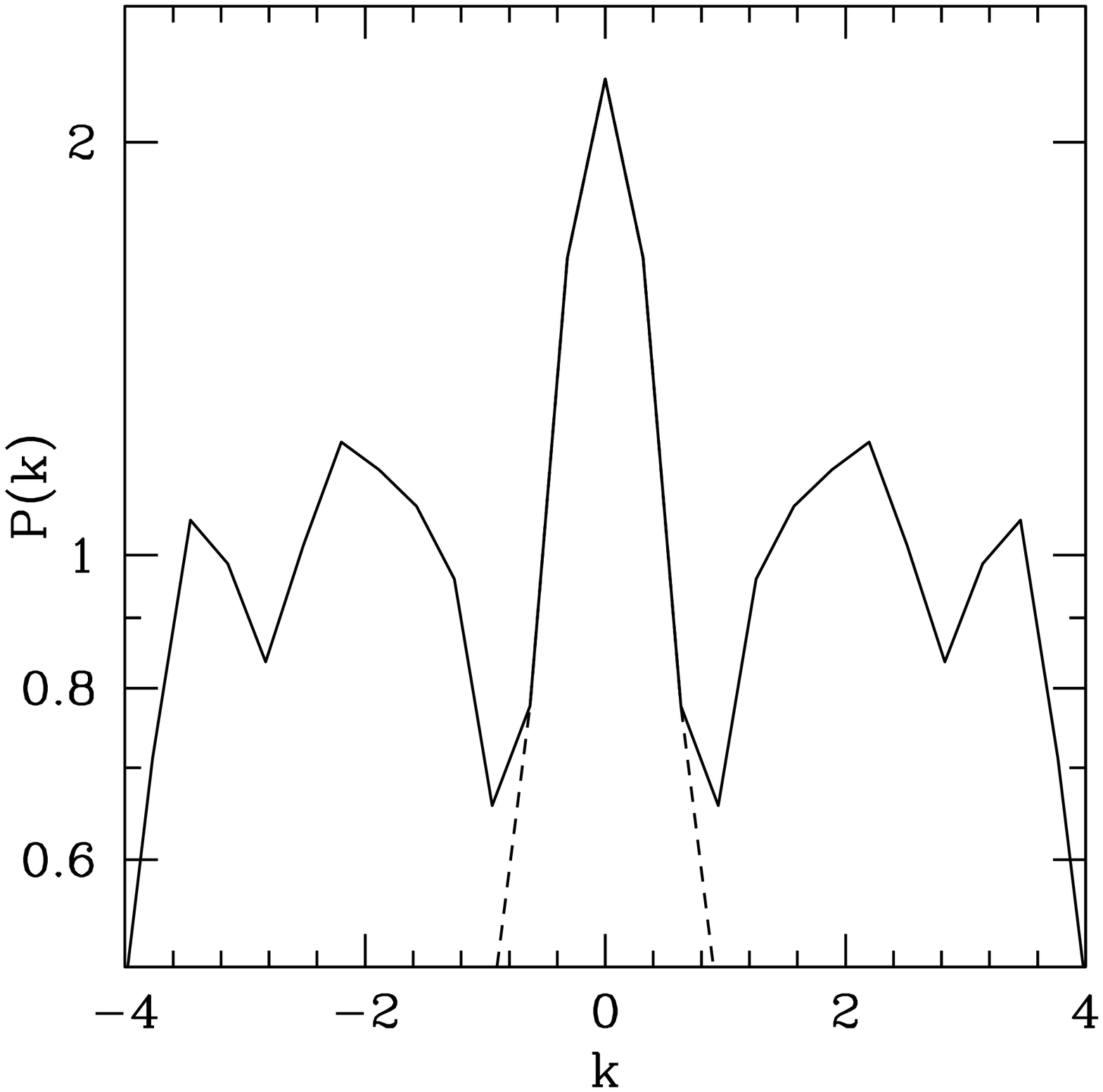}
\caption{The power spectrum of the SZ map. The dashed line indicates 
the location of the turning point. \label{fig:pow}}
\end{figure}

Let $P_{-}$ and $P_{+}$ be the real and corrupted power spectrum of 
$n_{cl}^{(0)}(\nu)$ respectively, then the Wiener filter in 
the Fourier $\nu_{k}$-space ($\nu_{k}$: the Fourier counter part 
of $\nu$) can be written as 
$W_{P}(\nu_{k}) = \frac{P_{-}(\nu_{k})}{P_{+}(\nu_{k})}$. 
One can easily see that $W_{P}(\nu_{k}) \approx 1$ where the error is 
negligible, and $W_{P}(\nu_{k}) \approx 0$ where the error is dominant.
Unfortunately, we cannot determine the exact functional form 
of $W_{P}(\nu_{k})$ since the only available quantity to us is 
$P_{+}(\nu_{k})$. Nevertheless, 
since the Wiener filter works in the least-square sense \cite{pre-etal92}, 
even a fairly reasonable approximation to $W_{P}(\nu_{k})$ can make it work 
quite well.   Figure \ref{fig:pow} plots $P_{+}(\nu_{k})$, showing 
there is a sharp boundary between the noise-dominant and 
negligible sections where $P_{+}(\nu_{k})$ has two distinct behaviors.   
Finding a turning point, $\nu_{k0}$, by eye, we approximated the 
Wiener filter by a step function such that 
$W_{P}(\nu_{k}) = \Theta(\vert\nu_{k}\vert < \nu_{k0})$, given the
asymptotic behaviors of $W_{P}(\nu_{k})$.   

Here is our recipe for the determination of the cluster luminosity 
function from the total SZ map: 
Construct $n_{cl}^{(0)}(\nu)$ on a one-dimensional discrete grid  
by counting the peak number density of the total SZ map 
and subtracting equation (3) from it. 
Calculate its power spectrum, $P_{+}(\nu_{k})$, by measuring the mean 
square amplitude of its Fourier transform. Plot $P_{+}(\nu_{k})$ to  
find the turning-point, $\nu_{k0}$. Approximate $W_{P}(\nu_{k})$ as a 
step-function such that 
$W_{P}(\nu_{k}) = \Theta(\vert\nu_{k}\vert < \nu_{k0})$. 
Convolve $n_{cl}^{(0)}(\nu)$ by $W_{P}(\nu_{k})$ and deconvolve it 
by the one-dimensional Gaussian distribution, $p(x)$. 

\begin{figure}
\plotone{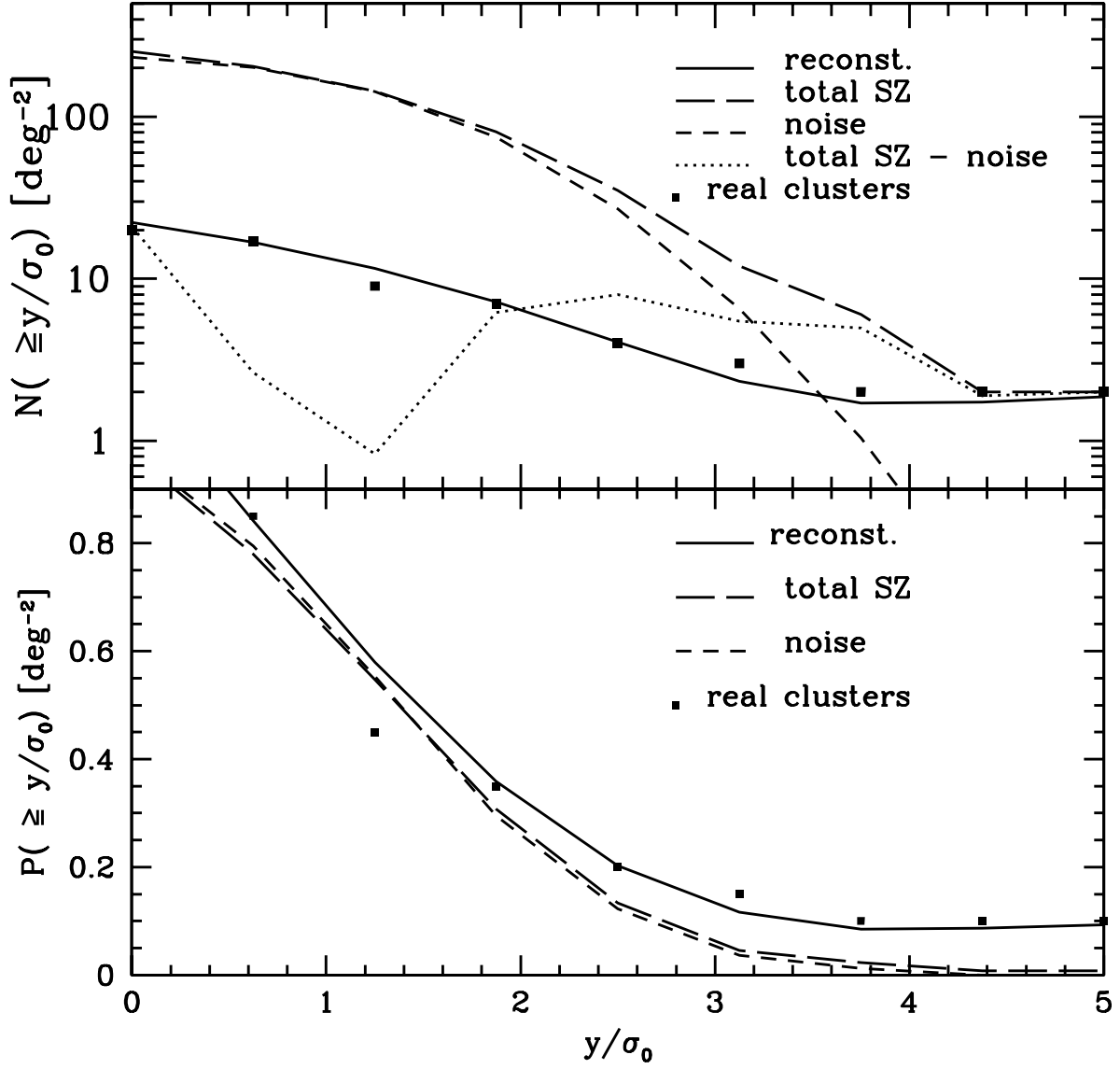}
\caption{{\it Upper}:The cumulative number density of local maxima as a 
function of the rescaled amplitude, $\nu$. The total number of cluster peaks 
($N_{cl}^{\rm tot}$) are assumed to be given as a prior. 
$N_{cl}^{\rm tot} = 20$ in this figure.
{\it Lower}: The cumulative probability distribution of local maxima as a 
function of the rescaled amplitude. Unlike the number density, the 
cluster probability distribution can be reconstructed without any prior.
\label{fig:dis}}
\end{figure}

In the upper panel of Figure \ref{fig:dis}, we show the cumulative 
cluster luminosity function $N_{cl}(\ge \nu)$ (solid line) 
reconstructed from the SZ map with the above recipe and compare 
it with the real distribution (square dots). We also plot the 
cumulative peak number density of the total SZ map (long-dashed line), 
the cumulative noise mean number density (dashed line), and the 
cumulative zeroth-order cluster luminosity function 
(dotted line) for comparison. 
Figure  \ref{fig:dis} reveals that the reconstructed cluster luminosity 
function is indeed in good agreement with the real distribution, 
especially in the medium range of $2 \le \nu \le 4$. Note also 
that in the low-$\nu$ section ($\nu \le 1$) noise dominates 
the SZ peaks while in the high-$\nu$ tail ($\nu \ge 4$) the total 
SZ peaks are mainly the cluster peaks as expected. 

When determining $n_{cl}(\nu)$, the total number of cluster peaks, 
$N_{cl}^{\rm tot}$, are assumed to be given as priors. In case that 
$N_{cl}^{\rm tot}$ is not available, we can still determine the 
probability density distribution of cluster peaks, 
$P_{cl}(\ge \nu) = N_{cl}(\ge \nu)/N_{cl}^{\rm tot}$. 
The lower panel of Figure \ref{fig:dis} plots the cumulative 
probability density distributions that can be determined using 
our analysis technique without any prior. 
Again, the real and the reconstructed distributions agree with each
other quite well. 
 
We tested our technique against different realizations of the SZ maps 
by varying the total number of cluster peaks and distribution shapes, 
and found it quite robust. 

\section{SUMMARY AND DISCUSSIONS}

We have developed a useful analysis technique to determine the cluster 
luminosity function efficiently, using AMiBA, a drift-scan
interferometric SZ survey, as a model experiment. We have simulated a 
total SZ map using the Monte-Carlo method, and counted the peak number 
density from it.  The total SZ map is constructed by combining a
Gaussian noise field with cluster sources. We noted that the peak number 
density of the SZ map at medium peak amplitude range has non-negligible 
contributions from the cluster peaks with noise-contamination effects 
included.   
 
To determine the cluster number density, i.e., the cluster luminosity 
function, first we have measured the zeroth-order 
cluster luminosity function by subtracting the available 
mean noise number density from the peak number density of the total 
SZ map. We have quantified the noise-contamination effects included in 
the zeroth-order approximation by a single Gaussian variable, 
and found that the cluster luminosity function can be expressed as 
the deconvolution of the zeroth-order approximation by the 
one-dimensional Gaussian distribution. 

We have stabilized the deconvolution process by convolving the 
zeroth-order approximation with a Wiener filter. 
The approximate functional form of the Wiener filter has been determined 
from the information of the power spectrum of the zeroth-order 
approximation. Finally by deconvolving the Wiener-filtered 
zeroth-order approximation of the cluster luminosity function 
we have determined the cluster luminosity function from the simulated 
SZ map. We have compared the reconstructed (cumulative) cluster 
luminosity function with the real one, and found good agreements 
between them, especially in the medium amplitude range where the 
conventional technique fails. 

The consequence of the statistical strategy presented here 
is that it can allow us to find the cumulative distribution of 
the sources even when the number of the sources occupy only 
small fraction of the total number of maximum peaks. 
We also expect this statistical strategy to be applied to the 
construction of the cluster mass function in weak gravitational 
lensing analysis. 

However, it is worth noting that the measurement accuracy of the 
cluster luminosity function may be improved on by improving the 
approximation accuracy of the Winer filter, and it is also worth 
noting that although our technique determines the cluster number 
density efficiently but it cannot select the cluster peaks from the SZ map.  
Furthermore, to examine the usefulness of our analysis technique 
in real practice, testing it against real SZ hydrodynamic simulations 
will be necessary. Our future work is in this direction.
 
\acknowledgments

We thank U. L. Pen for many helpful discussions on drift-scan method, 
and K. Yoshikawa for useful comments.  This work was supported by the 
research grants of the JSPS fellowship .

\clearpage
\clearpage

\newpage
\begin{thebibliography}{}
\bibitem[Bahcall \& Fan 1998]{bah-fan98}
Bahcall, N., \& Fan, X. 1998, \apj, 504, 1
\bibitem[Barbosa \etal\ 1996]{bar-etal96}
Barbosa, D. Bartlett, J. G., Blanchard, A., \& Oukbir, J. 1996, \aap, 314, 13
\bibitem[Benson \etal\ 2002]{ben-etal02}
Benson, A. J., Reichardt, C., \& Kamionkowski, M. 2002, \mnras, 331, 71  
\bibitem[Bond \& Efstathiou 1987]{bon-efs87}
Bond, J. R., \& Efstathious, G. 1987, \mnras,226, 655
\bibitem[Borgani \etal\ 1999]{bor-etal99}
Borgani, S., Rosati, P., Tozzi, P., \& Colin, N. 1999, \apj, 517, 40
\bibitem[Carlstrom \etal\ 1996]{car-etal96}
Carlstrom, J. E., Joy, M., \& Grego L. 1996, \apj, 456, 75
\bibitem[Diego \etal\ 2002]{die-etal02}
Diego, J. M., Martinez-Gonzalev, E., Sanz, J> L., Benitez, N., J. Silk 
2002, \mnras, 331, 556 
\bibitem[Fan \& Chiueh 2001]{fan-chi01}
Fan, Z., \& Chiueh, T. 2001, \apj, 550, 547 
\bibitem[Grego \etal\ 2001]{gre-etal01}
Grego, L., Carlstrom, J. E., Reese, E. D., Holder, G. P., 
Holzapfel, W. L., Joy, M. K., Mohr, J. J., \& Patel, S. 2001, 552, 2
\bibitem[Henry 1997]{hen97}
Henry, J. P. 1997, \apj, 489, L1
\bibitem[Holder \etal\ 2000]{hol-etal00}
Holder, G. P., Mohr,J. J., Carlstrom, J. E., Evrard, A. E., \& 
Leitch, E. M. 2000, \apj, 544, 629
\bibitem[Lo \etal\ 2000]{lo-etal00}
Lo, K. H., Chiueh, T. H., Martin, R. N., Ng, K. W., Liang, H., 
Pen, U. L, \&  Ma, C. P. 2000, preprint (astro-ph/0012282)
\bibitem[Longuet-Higgins 1957]{lon57}
Longuet-Higgins, M. S. 1957, Phil Trans Roy Soc London A, 249, 321 
\bibitem[Metzler 2002]{met02}
Metzler, C. A. 2002, preprint (astro-ph/9812295)
\bibitem[Molnar \etal\ 2002]{mol-etal02}
Molnar, S. M., birkinshaw, M.,\&  Mushotzky, R. F. 2002, \apj, 570, 1 
\bibitem[Pen \etal\ 2002]{pen-etal02}
Pen, U. L., Ng, K. W., Kesteven, M. J., \& Sault, B. 2002, preprint 
\bibitem[Press \etal\ 1992]{pre-etal92}
Press, W. H., Teukolsky, S. A., Vetterling, W. T., \& Flannery, B. P. 
1992, Numerical Recipes in Fortran (Univ. of Cambridge: New York)
\bibitem[Springel \etal\ 2001]{spr-etal01}
Springel, V., White, M., Hernquist, L. 2001, \apj, 549, 681 
\bibitem[Sunyaev \& Zel'dovich 1972]{sun-zel72}
Sunyaev, R. A., \& Zel'dovich, Y. B. 1972, Comm. Astrophys. Sp. Phys., 
4, 173 
\bibitem[Viana \& Liddle 1999]{via-lid99}
Viana, P. T. P., \& Liddle, A. R. 1999, 303, 535
\bibitem[White \etal\ 1999]{whi-etal99}
White, M., Carlstrom, J. E., Dragovan, M., \& Holzapfel, W. L. 1999, 
\apj, 514, 12
\bibitem[Zhang \etal\ 2002]{zha-etal02}
Zhang, P., Pen, U. L., \& Wang, B. 2002, preprint (astro-ph/0201375)
\end{thebibliography}
\end{document}